\documentstyle[pra,aps]{revtex}
\begin{document}
\input{epsf.tex}
\epsfverbosetrue

\title{ Stability of multi-parameter solitons: Asymptotic approach}
\author{Dmitry V. Skryabin 
\footnote{URL: http://cnqo.phys.strath.ac.uk/$\sim$dmitry}
}
\address{
Department of Physics and Applied Physics, John Anderson Building,\\ 
University of Strathclyde, 107 Rottenrow, Glasgow, G4 0NG, Scotland, UK}

\date{December 23, 1998}

\maketitle

\begin{abstract}
General  asymptotic approach to the stability problem
of  multi-parameter solitons in  Hamiltonian systems 
$i\partial E_n/\partial  z=\delta H/\delta E_n^*$ has been developed.
It has been shown that asymptotic study of the soliton
stability can be reduced to the calculation of a  certain sequence of the determinants,
where the famous determinant of the matrix consisting from the derivatives of the
system invariants with respect to the soliton parameters 
is just the first in the series. The presented approach 
gives first analytical  criterion  for the oscillatory instability and also predicts
novel stationary instability.
Higher order approximations allow to calculate corresponding  
eigenvalues  with  arbitrary accuracy.
\end{abstract}

PACS: 05.45.Yv,47.20.Ky,42.65.Tg,52.35.-g

Key words: soliton stability, Hamiltonian systems, codimension-2 bifurcation

\section{Introduction}
Solitary waves ('solitons') can appear  when an initial excitation  
applied to a medium is strong  enough to cause nonlinear  response. 
Formally solitons are solutions of some nonlinear partial differential equations
and their dynamics generally is a complex phenomenon, which can be described exactly 
only in  the very special {\em integrable} situations \cite{newell}.
The problems of soliton  stability  and instability induced dynamics 
in nonintegrable Hamiltonian models have paramount importance for understanding of a wide range of
physical phenomena covering such fields as propagation of electromagnetic, 
water and plasma waves, condensed matter physics  and classical 
field theory \cite{Malomed,Kuznetsov86,Kusmartsev,Makhankov,Pego92}. 
Several analytical approaches to  the  stability problem 
are known. For instance, in the nearly integrable situations the perturbation theory
based on the inverse scattering transform can be used \cite{newell,Malomed}.
Far from the integrable limit variety of  methods can be
applied. Among them  asymptotic stability theory \cite{Zakharov74a,Pelinovsky95}, 
method of adiabatically varying soliton parameters 
\cite{newell,Pelinovsky95,Pelinovsky,Buryak,Trillo98prl,Kaup,Grimshow},
Lyapunov \cite{Makhankov} and Evans \cite{Pego92}  methods.

Generally, stability of a solitary wave in a Hamiltonian model
can be lost due to bifurcations involving appearance of a positive eigenvalue
({\em stationary} instability) in the soliton spectrum or 
a pair of complex conjugate eigenvalues with positive real parts 
({\em oscillatory} instability) \cite{MacKay}.
Both types of these instabilities have been  
extensively studied in the different solitonic contexts proving their ubiquitousness
and fundamental importance, 
see Refs. \cite{Pego92,Zakharov74a,Pelinovsky95,Pelinovsky,Buryak,Torner,Skryabin98} 
and Refs. \cite{Pego92,Trillo98prl,Skryabin98,complex}, 
respectively, for the stationary and oscillatory instabilities.
In the most of the known cases the loss of stability is associated with
the collisions of the purely imaginary eigenvalues corresponding
to the so called internal modes \cite{int_mode} of the soliton spectrum 
(see \cite{Pego92} for interesting exceptions).

Applying the above mentioned  methods it was shown that in many cases 
a threshold of  stationary instability of  multi-parameter solitons 
 is given by the zero of the determinant of the Jacobi matrix 
$J_{ij}=\partial_{\kappa_j}Q_i$,
where $\kappa_j$ are the soliton parameters and $Q_i$ are the associated motion integrals 
\cite{Kuznetsov86,Kusmartsev,Makhankov,Pego92,Zakharov74a,Pelinovsky95,Pelinovsky,Buryak,Trillo98prl,Torner}. 
The  condition $det(J_{ij})=0$ is, in fact, the compatibility condition of the problem arising
in the leading (zero) order of the asymptotic solution of the eigenvalue problem governing
stability of the soliton \cite{Pelinovsky95,Buryak,Torner}. 
To find expressions for the eigenvalues it is necessary to proceed further and solve 
problems arising in  higher (at least first) orders. 
Up to now this was done only for the specific class of  model equations
having single parameter soliton families \cite{Zakharov74a,Pelinovsky95}. 
For stationary bifurcations of  two-parameter solitons adiabatic method 
 has been applied in Refs. \cite{Buryak,Trillo98prl,Grimshow}. 
Linear approximation of this method actually gives an expression for eigenvalues,
for more details see Section IV. 
However, all known developments of this method fail to give criterion  
indicating transition to the oscillatory instability, i.e. instability with complex eigenvalues.
It is also difficult  to extend this method beyond its first order because of the 
rather involved calculations.

{\em The  purpose of this work is to formulate a general asymptotic approach 
to  stability of  multi-parameter solitons in Hamiltonian models, to show how it 
can be used to find expressions for the instability growth rates with arbitrary accuracy 
and to formulate criterion for the  oscillatory instability  of  solitons.}

\section{Model equations and symmetries}

We will consider Hamiltonian  equations  in the form 
\begin{equation}
i\frac{\partial E_n}{\partial z}=\frac{\delta H}{\delta E_n^*},~n=1,2\dots N,
\label{eq1}\end{equation}
which describes a wide range of physical phenomena
related with self-action and interaction of  slowly varying 
wave envelopes in a variety of nonlinear media
\cite{Kuznetsov86,Pelinovsky,Buryak,Trillo98prl,Kaup,Torner,Skryabin98,complex},
for general review of the Hamiltonian formalism see \cite{usp}.
Here $E_n$ are  complex fields,
$z$  the propagation direction of the interacting waves, 
$x$  the coordinate characterizing dispersion or diffraction, 
$H=H(\partial_xE_n,E_n,\partial_xE_n^*,E_n^*)$ is the Hamiltonian and $^*$ means complex
conjugation. We will assume that $H$ is invariant with respect to a set of $(L-1)$ 
phase transformations:
\begin{equation}
E_n\to E_n\exp(i\gamma_{nl}\phi_l),~l=1,2,\dots (L-1),
\label{eq2}\end{equation}
$\phi_l$ are  arbitrary real phases and $\gamma_{nl}$ are some constants.
Because $H$ does not depend on $x$ explicitly, Eqs. (\ref{eq1}) are also 
invariant with respect to  arbitrary translations along $x$: 
\begin{equation}
E_n(x,z)\to E_n(x-x_0,z).
\label{eq3}\end{equation} 

Symmetry properties (\ref{eq2}), (\ref{eq3}) together with Hamiltonian nature of our problem
imply presence of $L$ conserved quantities, see, e.g., \cite{form}, which are  the 
$(L-1)$ energy invariants
\begin{equation}
Q_l=\int dx\sum_{n=1}^{N}\gamma_{nl}|E_n|^2,~l=1,2,\dots (L-1),
\label{eq4}\end{equation}
and the momentum
\begin{equation}
Q_L=\frac{1}{2i}\int dx\sum_{n=1}^{N}(E_n^*\partial_xE_n-E_n\partial_xE_n^*)
\label{eq5}.\end{equation}

Another important consequence of the invariances (\ref{eq2}), (\ref{eq3}) is
that a certain class of solutions of Eqs. (\ref{eq1}) can be sought in a form when $x_0$ 
and $\phi_l$ are   linear functions of $z$, i.e. $x_0=\kappa_Lz$ and $\phi_l=\kappa_lz$, then  
\begin{equation}
E_n(x,z)=a_n(x-\kappa_Lz)\exp(i\sum_{l=1}^{L-1}\gamma_{nl}\kappa_l z)
\label{eq6},\end{equation}
where $\{\kappa_l\}_{l=1}^{L-1}$ and $\kappa_L$   are  real parameters characterizing, respectively, 
phase  velocities of the interacting waves
and  the soliton group velocity.
Functions $a_n(\tau)$ obey  a system of the ordinary differential equations
\begin{equation}
(i\kappa_L\partial_{\tau}+\alpha_n)a_n=-\frac{\delta H_a}{\delta a_n^*}
\label{eq7},\end{equation}
where  $H_a\equiv H(\partial_{\tau}a_n,a_n,\partial_{\tau}a_n^*,a_n^*)$, 
$\tau=x-\kappa_Lz$ and $\alpha_n=\sum_{l=1}^{L-1}\gamma_{nl}\kappa_l$.
We assume now that in a certain domain of the parameter space $(\kappa_1,\kappa_2,\dots\kappa_L)$  
Eqs. (\ref{eq7}) have a family of the solitary solutions such that $|a_n|\to 0$ for $\tau\to\pm\infty$.

\section{Asymptotic stability analysis}

To study stability of the solitons  we seek solutions
of  Eqs. (\ref{eq1}) in the form
\begin{equation}
E_n=(a_n(\tau)+\varepsilon_n(\tau,z))\exp(i\sum_{l=1}^{L-1}\gamma_{nl}\kappa_l z),
\label{eq8}\end{equation}
where $\varepsilon_n(\tau,z)$ are  small complex perturbations.
Linearizing  Eqs. (\ref{eq1}) and assuming that $\varepsilon_n(\tau,z)=\xi_n(\tau)e^{\lambda z}$,
$\varepsilon_n^*(\tau,z)=\xi_{n+N}(\tau)e^{\lambda z}$ we get the following nonselfadjoint 
eigenvalue problem (EVP) 
\begin{equation}
i\lambda\vec\xi=\hat{\cal L}\vec\xi\equiv
\left(\begin{array}{cc}
\hat S & \hat R\\ -\hat R^* & -\hat S^*
\end{array}\right) \vec\xi
\label{eq9},\end{equation}
where $\vec\xi=(\xi_1,\dots\xi_N,\xi_{N+1}\dots\xi_{2N})^T$, and $\hat R$, $\hat S$
are $N\times N$ matrix operators with elements given by
$$\hat s_{nl}=\delta_{nl}(\alpha_n+i\kappa_L\partial_{\tau})+\frac{\delta^2 H_a}{\delta a_n^*\delta a_l},~
\hat r_{nl}=\frac{\delta^2 H_a}{\delta a_n^*\delta a_l^*},$$
here $\delta_{nl}$ is the Kroneker symbol. Note, that the operator $\hat S$ is a selfadjoint one, 
i.e. $\hat S=\hat S^{\dagger}$, and $\hat R$ is a symmetric operator, i.e. $\hat R=\hat R^T$.

To solve EVP (\ref{eq9}) we  apply  the asymptotic approach, which
 relies on  expansion of the unknown eigenvector $\vec \xi$
into an asymptotic series near either neutral eigenmodes 
 \cite{Zakharov74a,Pelinovsky95}, i.e. zero-eigenvalue modes,  of the operator $\hat{\cal L}$,
or modes of continuum \cite{int_mode}, or both of them \cite{Kaup}. 
The neutral modes can  be generated by  infinitesimal
variations of the free parameters of the soliton  and thus  always
 be presented as  explicit functions of the soliton solution. 
 At the same time continuum eigenmodes are explicitly known
in the very rare, normally in integrable, situations \cite{Kaup,int_mode}. 
This is an important fact which makes the asymptotic expansion near the neutral modes
by the very practical tool of the stability theory. However, as any approximate method,
it has a certain limitation. Namely, it  describes only eigenvalues $\lambda$
corresponding to a specific class of the perturbations which in the zero  approximation
can be expressed as a linear superposition of the neutral eigenmodes. 
Thus, generally speaking, on the basis of this approach one can get only 
{\em sufficient}  conditions for soliton
instability or, in other words, {\em necessary} conditions for  soliton stability.
Therefore presence of other instabilities which can be captured only numerically
can always be expected \cite{Skryabin98}.

By  infinitesimal variation of $\phi_l$ and $x_0$  it can be shown that 
\begin{eqnarray}
\nonumber && \vec u_l=(\gamma_{1l}a_1,\dots\gamma_{Nl}a_N,-\gamma_{1l}a_1^*,
\dots -\gamma_{Nl}a_N^*)^T,~
\vec u_{L}=\frac{\partial \vec a}{\partial\tau},\\ 
\nonumber && \vec a\equiv(a_1,\dots a_N,a_1^*,\dots a_N^* )^T,~l=1,\dots (L-1)
\end{eqnarray}
are neutral  modes of $\hat{\cal L}$, i.e. $\hat{\cal L}\vec u_{l}=0$ 
$(l=1,\dots L)$. $\hat{\cal L}$ also has $L$ associated vectors 
$\vec U_l=\partial\vec a/\partial\kappa_l$  such that
$\hat{\cal L}\vec U_{l}=-\vec u_{l},~l=1,\dots L$.

It is straightforward to see that any solution of EVP (\ref{eq9})
 must obey $L$ solvability conditions
\begin{equation}
\langle\vec w_l|\lambda\vec\xi\rangle=0,~l=1,\dots L,
\label{eq13}\end{equation}
where $\langle\vec y|\vec z\rangle=\sum_{i=1}^{2N}\int dxy_i^*z_i$ and
$\vec w_{l}$ are the neutral modes of the operator $\hat{\cal L}^{\dagger}$, 
$\hat{\cal L}^{\dagger}\vec w_{l}=0$,
\begin{eqnarray}
\nonumber && \vec w_l=(\gamma_{1l}a_1,\dots\gamma_{Nl}a_N,\gamma_{1l}a_1^*,\dots \gamma_{Nl}a_N^*)^T,
~\vec w_L=i\frac{\partial\vec b}{\partial\tau},\\
\nonumber && \vec b=(-a_1,\dots -a_N,a_1^*,\dots a_N^*)^T,~l=1,\dots (L-1).
\end{eqnarray} 
Associated vectors of $\hat{\cal L}^{\dagger}$ are $\vec W_l=\partial\vec b/\partial\kappa_l$ 
and they obey $\hat{\cal L}^{\dagger}\vec W_{l}=-\vec w_{l},~l=1,\dots L$.

Close to instability threshold it is naturally to assume that $|\lambda|\sim\epsilon\ll 1$.
As it was already discussed above we will consider a special class of the perturbations 
which   in the leading approximation can be presented as a linear combination 
of the neutral modes. Therefore we seek an asymptotic solution of EVP
(\ref{eq9}) in the following form
\begin{equation}
 \vec\xi=\sum_{m=0}^{\infty}\epsilon^m\vec\xi_m(x),
~\vec\xi_0=\sum_{l=1}^{L}C_{l}\vec u_{l}
\label{eq11}\end{equation} 
where constants $C_{l}$ and vector-functions $\vec\xi_{m>0}$ have to be defined.
Here and below $l=1,2,\dots L$.
Substitution (\ref{eq11}) into EVP (\ref{eq9}) gives a recurrent system of 
equations for $\vec\xi_m$
\begin{equation}
\vec\xi_{m>0}=\left[\frac{i\lambda}{\epsilon}\hat{\cal L}^{-1}\right]^m\vec\xi_0.
\label{eq12}\end{equation}

Substituting (\ref{eq11}), (\ref{eq12})  into conditions (\ref{eq13})
 one will find the homogeneous system of the $L$ linear algebraic equations 
\begin{equation}
\lambda^2\Big\langle \vec w_{l}\Big|
\sum_{m=0}^{\infty}(-\lambda^2)^m\hat{\cal L}^{-2m}
\sum_{l=1}^{L}C_l\vec U_l\Big\rangle=0
\label{eq15}\end{equation}
for $L$ unknown constants $C_{l}$.
System (\ref{eq15}) has a nontrivial solution providing that the 
corresponding determinant is equal to zero. This determinant is
an infinite-order polynomial with respect to $\lambda^2$, which, 
in fact, is the asymptotic expansion of  Evans function \cite{Pego92}. 
Zeros of this polynomial define the spectrum of the solitary wave linked with 
the chosen class of the perturbations. Thus the equation specifying eigenvalues $\lambda$ is
\begin{equation}
 \lambda^{2L}\sum_{j=0}^{\infty}(-\lambda^2)^jD_j=0,
\label{eq16}\end{equation}
where $D_j$ are the real constants. 
Eq. (\ref{eq16})  always has zero root of the $2L$-order. 
It indicates that each of the zero eigenvalues corresponding 
to the neutral modes $\vec u_{l}$ is doubly degenerate one. This degeneracy
originates from the presence of the associated vectors $\vec U_{l}$.

To write the explicit expressions for $D_j$ it will be convenient 
to introduce vectors 
$\vec{\cal M}_l^{(m)}=({\cal M}_{l1}^{(m)}\dots{\cal M}_{lL}^{(m)})$, where,
$${\cal M}_{ll^{'}}^{(m)}=\langle\vec w_l|\hat{\cal L}^{-2m}\vec U_{l^{'}}\rangle, 
~m=0,1,\dots\infty.$$
Now each $D_j$ can be presented as 
\begin{equation}
D_j=\sum_{m_1+\dots m_L=j}{\cal D}(\vec{\cal M}_1^{(m_1)},\dots\vec{\cal M}_L^{(m_L)}),
\label{eq16a}\end{equation}
where ${\cal D}(\vec{\cal M}_1^{(m_1)},\dots\vec{\cal M}_L^{(m_L)})$ 
is the determinant of the $L\times L$ matrix consisting of the
rows  $\vec{\cal M}_l^{(m_l)}$ and the sum is taken over all such combinations
of $(m_1,\dots m_L)$ that $\sum_{l=1}^Lm_l=j$. 
${\cal M}^{(0)}_{ll^{'}}$ can be readily expressed via
derivatives of the conserved quantities with respect to the soliton parameters:
\begin{equation}
\nonumber {\cal M}^{(0)}_{ll^{'}}=\frac{\partial Q_l}{\partial\kappa_{l^{'}}},
\end{equation}
and practical calculation of ${\cal M}^{(m)}_{ll^{'}}$ for $m>0$ can be simplified:
${\cal M}^{(m)}_{ll^{'}}=-\langle\vec W_l|\hat{\cal L}^{(1-2m)}\vec U_{l^{'}}\rangle.$
Note, that in most of the cases  solitary solution itself can be found only numerically
using any of the well established methods for solving  the nonlinear {ode}'s.
Recurrent calculations of $\hat{\cal L}^{(1-2m)}\vec U_{l}$ can be readily reduced
to the numerically even simpler problem of solving of the linear inhomogeneous ode's.

\begin{figure}
\setlength{\epsfxsize}{7.0cm}
\centerline{\epsfbox{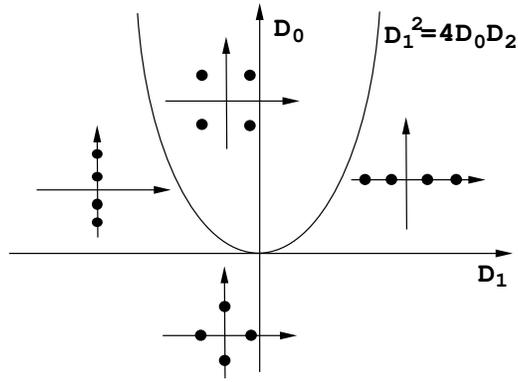}}
\vspace{3mm}
\caption{Soliton bifurcation diagram in the neighbourhood of the point $D_0=D_1=0$ 
for $D_2>0$. Insets show $(Re\lambda,Im\lambda)$-plane with horizontal/vertical axes
corresponding to $Re\lambda/Im\lambda$ and dots
marking soliton eigenvalues described by  Eq. (\ref{eq20}).}
\label{fig1}
\end{figure}

Because $|\lambda|$ was assumed to be small, Eq. (\ref{eq16}) has an asymptotic character.
Therefore to make it work some additional assumptions must be made about orders of $D_j$. 
If these assumptions are satisfied then 
Eq. (\ref{eq16})  describes correctly the soliton spectrum and  predicts bifurcations of the soliton.
The corresponding eigenvalues can be found using Eq. (\ref{eq16}) with any degree of accuracy.
For example, let us assume that $D_0\sim\epsilon^2$ and  $D_{j>0}\sim O(1)$.
Then, presenting $\lambda^2$ as
\begin{equation}
\lambda^2=\epsilon^{2}\sum_{j=0}^{\infty}\zeta_j,~\zeta_j\sim\epsilon^{2j},
\label{eq17}\end{equation}
in the first order Eq. (\ref{eq16}) gives a linear equation for $\zeta_0$, 
\begin{equation}
D_0-\epsilon^{2}\zeta_0D_1=0,\label{eq18}\end{equation}
which indicates a threshold of the stationary bifurcation at $D_0=0$.
This is precisely the condition  $det(J_{ij})=0$ discussed in the introduction.
Continuing to the next order one obtains
\begin{equation}
\lambda^2=\frac{D_0}{D_1}\left(1-\frac{D_0D_2}{D_1^2}+O(\epsilon^4)\right).
\label{eq19}\end{equation}
If $D_1\sim\epsilon^2$ then the asymptotic expression (\ref{eq19}) fails. 
 To have a balanced equation for $\zeta_0$, we must now assume that $D_0\sim\epsilon^4$. 
However, in such a case the Eq. \ref{eq18} for $\zeta_0$  
changes from  linear  to  quadratic:
\begin{equation}
D_0-\epsilon^{2}\zeta_0D_1+\epsilon^4\zeta_0^2D_2=0.
\label{eq20}\end{equation}
Eq. (\ref{eq20}) gives two threshold conditions $D_0=0$ and $D_1^2=4D_0D_2$, 
see Fig. \ref{fig1}. The latter condition indicates onset of 
the oscillatory instability for 
\begin{equation}
D_1^2<4D_0D_2.
\label{eq20a}\end{equation}
 Thus we have formulated analytic criterion for the oscillatory instability.
It is also clear that the point $D_{0,1}=0$  is a source for the novel  stationary instability,
see rightmost region $D_1^2>4D_0D_2$, $D_1>0$ in Fig. \ref{fig1}, where an eigenvalue
which is positive throughout this region can not be predicted by Eq. (\ref{eq19}).

It follows by recurrence  that if $D_{j^{'}>0}\sim\epsilon^2$
 then to have a balanced equation for $\zeta_0$ we must assume that 
$D_{j<j^{'}}\sim\epsilon^{2(1+j^{'})}$.
In other words asymptotic expansion near the neutral modes can only describe  the 
soliton spectrum in  regions of the parameter space which  are close to 
codimension-$(j^{'}+1)$ bifurcation. If  $j^{'}=0$ then only one condition 
must be satisfied  and our asymptotic approach predicts presence of either
two purely imaginary or two purely real eigenvalues, which can collide at zero.
If $j^{'}=1$ then two conditions  must be satisfied 
and the asymptotic approach predicts presence of two pairs of  eigenvalues which
can be real, imaginary or complex. In this situation the soliton becomes oscillation 
unstable  providing that two pairs of imaginary eigenvalues collided.
For each further  $j^{'}$ two new eigenvalues come into play.

\section{Discussion}
General formulae (\ref{eq16}),(\ref{eq16a})  giving  soliton eigenvalues 
with any degree of accuracy and criterion for the oscillatory instability (\ref{eq20a})
are main novel results of this work. At the same time expressions for the eigenvalues
near the stationary instability threshold, analogs of the formula $\lambda^2=D_0/D_1+...$,
 have been earlier obtained  in a number of papers.
It is instructive  now to give explicit expressions for  $D_j$
in the two simplest situations of  one- and two-parameter solitons and to compare them
with previously reported results.
For the one parameter solitons: $D_0=\partial_{\kappa_1}Q_1$, 
$D_1=-\langle\vec W_1|\hat{\cal L}^{-1}\vec U_1\rangle$,
$D_2=-\langle\vec W_1|\hat{\cal L}^{-3}\vec U_1\rangle$.
Using these formulae one can show that in the case  when $D_1\sim O(1)$ 
the first term in Eq. (\ref{eq19})
gives the same expression for $\lambda^2$ which was obtained in Refs. 
\cite{Zakharov74a,Pelinovsky95,Pelinovsky}, where generalised Nonlinear Shr\"odinger
equation \cite{Zakharov74a,Pelinovsky} and equations describing propagation in quadratically 
nonlinear media  \cite{Pelinovsky95} have been investigated. 
If $D_1D_2>0$  then it can be concluded that the 
second term in Eq. (\ref{eq19}) indicates saturation of the  growth rate 
when the distance from the instability threshold, $D_0=0$, growthes, which
agrees with  numerical results \cite{Pelinovsky95,Skryabin98}. 

For the two-parameter solitons:
\begin{eqnarray}
&& D_0=\left|
\begin{array}{cc}
\frac{\partial Q_1}{\partial\kappa_1} &\frac{\partial Q_1}{\partial\kappa_2}\\
\frac{\partial Q_2}{\partial \kappa_1} &\frac{\partial Q_2}{\partial\kappa_2}
\end{array}\right|,
\label{eq22a}\\
&& D_1=\left|
\begin{array}{cc}
\frac{\partial Q_1}{\partial\kappa_1} &\frac{\partial Q_1}{\partial\kappa_2}\\
{\cal M}_{21}^{(1)} &{\cal M}_{22}^{(1)}
\end{array}\right|+
\left|
\begin{array}{cc}  
{\cal M}_{11}^{(1)} &{\cal M}_{12}^{(1)}\\
\frac{\partial Q_2}{\partial \kappa_1} &\frac{\partial Q_2}{\partial\kappa_2}
\end{array}\right|,
\label{eq22b}\\
&& D_2=
\left|
\begin{array}{cc}
{\cal M}_{11}^{(1)} &{\cal M}_{12}^{(1)}\\
{\cal M}_{21}^{(1)} &{\cal M}_{22}^{(1)}
\end{array}\right|+
\left|
\begin{array}{cc}
\frac{\partial Q_1}{\partial\kappa_1} &\frac{\partial Q_1}{\partial\kappa_2}\\
{\cal M}_{21}^{(2)} &{\cal M}_{22}^{(2)}
\end{array}\right|+
\left|
\begin{array}{cc}  
{\cal M}_{11}^{(2)} &{\cal M}_{12}^{(2)}\\
\frac{\partial Q_2}{\partial \kappa_1} &\frac{\partial Q_2}{\partial\kappa_2}
\end{array}\right|\label{eq22c}.
\end{eqnarray}
The threshold condition $D_0=0$ has been previously found for  two-parameter solitons
in different physical contexts \cite{Buryak,Trillo98prl}. However derivation of an accurate 
expression for the soliton eigenvalues near this threshold has remained a controvertial problem.
Indeed, comparing  eigenvalues given by  Eqs. (\ref{eq19}), (\ref{eq22a}), (\ref{eq22b}),(\ref{eq22c})
and eigenvalues which can be calculated from  
the ordinary differential equations for soliton parameters presented in 
\cite{Buryak,Trillo98prl} one will discover  that results are slightly different \cite{comment}. 
It has also been argued \cite{Buryak} that the sign of 
$({\cal M}_{11}^{(1)}{\cal M}_{22}^{(1)}-{\cal M}_{12}^{(1)}{\cal M}_{21}^{(1)})$, which is  
the first term in  Eq. (\ref{eq22c}), plays an important role in  stability of two-parameter solitons.
However  Eqs. (\ref{eq19}),(\ref{eq22a}),(\ref{eq22b}),(\ref{eq22c})
apparently conflict with this finding.

Among open problems I would like to mention derivation of  finite-dimensional
normal forms describing dynamical evolution of the soliton parameters 
near the oscillatory instability threshold.
A guideline for this work can be  theory of G. Iooss \cite{iooss,alan} 
for the normal forms of the reversible ordinary differential equations \cite{alan}
in   vicinity of the codimension-2 bifurcation,
wich is an equivalent of the our point $D_0=D_1=0$. 
The simplest case of the codimension-1 stationary instability, $D_0=0$,   
has only one homoclinic orbit separating regions of  the periodic oscillations
from the  spreading or collapse \cite{Pelinovsky,Grimshow}.
The vicinity of the codimension-2 point can contain the very reach dynamics,
including multiple homoclinic orbits and stochastic regimes.

\section{Summary}

General form of the asymptotic approach to  stability problem 
of  multi-parameter solitons in  Hamiltonian systems has been developed.
It has been shown that the asymptotic study of the soliton
stability reduces to the calculation of a certain sequence of  determinants,
where the famous determinant of the matrix consisting from the derivatives of the
system invariants with respect to the soliton parameters \cite{Kusmartsev,Makhankov,Buryak} 
is just the first in the series. 
Knowledge of these determinants  allows to calculate   
eigenvalues governing soliton instability with  arbitrary accuracy. 
The most important consequence is that  the presented approach gives first  analytic
criterion for the oscillatory instability of solitons in Hamiltonian systems.

\section{Acknowledgment}
Author acknowledges useful discussions with  W.J. Firth and D.E. Pelinovsky and
financial  support from the Royal Society of Edinburgh and British Petroleum.



\vspace{-2mm}


\begin{references}
\bibitem{comment}
E.g. reading page 86  after Eq. (10) of Ref. \cite{Trillo98prl}  one can find
that $\lambda^2=-J/(m_v\partial_{\Delta}Q_0)$, where $J$ is exactly 
$D_0$ given by Eq. (\ref{eq22a}) and the denominator can be 
rewritten in our notations as
$-\frac{(\partial_{\kappa_2}Q_1)^2}{\partial_{\kappa_1}Q_1}{\cal M}_{11}^{(1)}-
{\cal M}_{22}^{(1)}\partial_{\kappa_1}Q_1+{\cal M}_{12}^{(1)}\partial_{\kappa_1}Q_2+
{\cal M}_{21}^{(1)}\partial_{\kappa_2}Q_1$. One should expect that 
the last expression is equal to $-D_1$, but it is not, due to its first term.
Source of this disagreement seems to be hidden in the fact that authors of 
Refs. \cite{Buryak,Trillo98prl} develop adiabatic theory of the order $\epsilon^2$
using the ratio between  time  derivatives of the soliton parameters 
obtained from the solvability
condition valid only in the order $\epsilon$, see text after Eq. (9) 
in \cite{Trillo98prl} 
and Eqs. (7), (10) in \cite{Buryak}.
\end{references}
\end{document}